\documentclass{article}
\usepackage[subpreambles=true]{standalone}
\usepackage{import}
\usepackage{amsmath}
\usepackage{calligra}
\usepackage{longtable}
\usepackage{authblk}

\input epsf
\usepackage{lipsum}

\begin{document}
\title{BINGO-ABDUS: a radiotelescope to unveil the dark sector of the Universe}
\author[1]{\small Elcio Abdalla
\thanks{eabdalla@if.usp.br}}
\author[2]{ Alessandro Marins
\thanks{alessandrormarins@gmail.com}}
\author[1,3]{ Filipe Abdalla\thanks{filipe.abdalla@gmail.com}}
\author[1]{Jordany Vieira}
\author[1]{Lucas Formigari}
\author[4]{Amilcar R. Queiroz}
\author[5,6]{Bin Wang}
\author[4]{Luciano Barosi}
\author[6,7]{Thyrso Villela}
\author[6]{Carlos A. Wuensche}
\author[9,10,11]{Chang Feng}
\author[4]{Edmar Gurjao}
\author[12]{Ricardo Landim}
\author[6]{Camila P. Novaes}
\author[4]{Joao R.L. Santos}
\author[13]{Jiajung Zhang}

\affil[1]{Departamento de Física Geral, Instituto de Física, Universidade de São Paulo, São Paulo, Brazil}
\affil[2]{Department of Astronomy, School of Physical Science, University of Science and Technology of China, Hefei, China}
\affil[3]{Department of Physics and Astronomy, University College London,  UK}
\affil[4]{Unidade Acad\^emica de F\'{i}sica, Univ. Federal de Campina Grande, R. Apr\'{i}gio Veloso, 58429-900 - Campina Grande, Brazil}
\affil[5]{School of Aeronautics and Astronautics, Shanghai Jiao Tong University, Shanghai 200240, China}
\affil[6]{Center for Gravitation and Cosmology, Yangzhou University,  China}
\affil[7]{Instituto Nacional de Pesquisas Espaciais, Divis\~ao de Astrof\'isica, S\~ao Jos\'e dos Campos, SP, Brazil}
\affil[8]{Instituto de F\'{i}sica, Universidade de Bras\'{i}lia, - Bras\'{i}lia, DF, Brazil}
\affil[9]{Department of Astronomy, School of Physical Science, University of Science and Technology of China, Hefei, China}
\affil[10]{CAS Key Laboratory for Research in Galaxies and Cosmology, University of Science and Technology of China, Hefei, China}
\affil[11]{School of Astronomy and Space Science, University of Science and Technology of China, Hefei, China}
\affil[12]{Technische Universit\"at M\"unchen, Physik-Department T70, James-Franck-Strasse 1, 85748, Garching, Germany}
\affil[13]{Shanghai Astronomical Observatory, Chinese Academy of Sciences,  China}
\maketitle

\begin{abstract}

The \emph{Baryon Acoustic Oscillations from Integrated Neutral Gas Observations} (BINGO) telescope is an international collaboration, led by Brazil and China, aiming to explore the Universe's history through integrated post-reionization 21cm signals and fast radio emissions. It is concentrated on the important period for understanding the recent Universe dynamics that have shed light on the existence of the Dark Sector. In its first phase, BINGO will explore the southern celestial Hemisphere to observe neutral hydrogen signals from when the Universe was between 8.969 - 12.112 billion years old\footnote{These age values are from the cosmology provided by PLANCK satellite \cite{planck2018_VI} assuming a flat-$\Lambda$CDM with $H_0 = 67.36$ km/s/Mpc, $\Omega_m = 0.3153$, and $\Omega_{\Lambda} = 0.6847$.} in an angular resolution of 40 arcmins, and for the next phases reaching far epochs when the Universe was 3.116 billion years old up to 27 arcmin resolution which is poor for identifying individual sources, however important for mapping out the Universe in large scales using the \emph{Intensity Mapping} observational method. For identifying individually fast radio sources, the \emph{Advanced Bingo Dark Universe Studies} (ABDUS) project has been proposed and developed and will combine the current BINGO construction with the main single-dish telescope and stations of phased-array and outrigger, which will reach out up to 2.5 arcsec resolution to pinpoint individual radio sources as Fast Radio Bursts by interferometry.\\

This work has been presented in Syros, Greece, in September 2022.
\end{abstract}
\section{Introduction} \label{sec:intro}

From a very early age, curious young people and future explorers are confronted with questions that still have no answers. An experienced explorer recognizes that not having all the answers is acceptable as long as the fire of curiosity continues. Open questions in human understanding are engines for scientific and technological development. In the 1930s, one of the most current meaningful scientific milestones comes up through ignorance about matter distribution and about the behavior of 
galaxies as a consequence of the existence of a Dark Sector, as we know a posteriori \cite{darkmatter_zwicky_2009_first_publ_ENGversion, darkmatter_zwicky_1937}. Dark Matter (DM) emerges with a generic name for a class of candidates for filling in the gap raised in the unseen matter quantity to describe the observed universe. On the other hand, the decade of the 1990s started the running to understand the nature of the Universe's accelerated expansion. Dark Energy (DE) is the name that condenses all fluid candidates to explain the Universe's accelerated behavior within Einstein's gravity. DM and DE together are complex challenges to the current description from elementary known principles since no natural representative arises from Particle Standard Model. DM and DE form the named \emph{Dark Sector} \cite{AM2020}, and its nature drives several new projects in current physics and astronomy studies.

However, in contrast to completely new proposals, an explanation of the above-unknown components needs to follow the usual thread of the already-known Universe's history with high accuracy. Or else provide a highly well-grounded new physics. For these, many distinct observational methods and/or tools are used aiming to elucidate more and more the components from different perspectives. The most promising way to explore the Universe's history and its contents is from the emission lines of neutral hydrogen (HI), the so-called 21cm signal, also known as hyperfine structure. The former name comes from the size of the wavelength of the emission. To be more precise, atomic neutral hydrogen emits a characteristic signal from the hyperfine splitting of the 1S ground state. Such signal has a wavelength of 21cm, corresponding to the frequency $\nu_{10}\ =$ 1420 MHz in the rest frame. Even with a very low transition probability ($\sim 10^{-15}$ s), in an astrophysical context, HI clouds contain a large amount of HI that can be excited - either by radiative transition or by collision - and emit radiation \cite{field1958}.

Although 21cm emissions have been used in astronomy for a long time, their use in astrophysics and cosmology is quite recent. Considering as precursor works of the 21cm for astronomy those of George Field  \cite{field1958,field1959,field1958b}, and even though the first estimates of the signal are around $\sim$ 1990-2000, the effective interest in emission in cosmology and astrophysics from the most primordial periods of the Universe begins with the work \cite{sunyaev1972} suggesting the use of 21cm emission lines to search for spectral or spatial signatures of structures at high redshifts. The first searches for the 21cm signal in this context took place around the 1980s, but it really gained momentum with the construction of the \emph{Giant Metrewave Radio Telescope} (GMRT) \cite{swarup1991}, in India in the late 1990s. Since then, several projects have been developed or are under development aiming to use this signal, and due to the HI being a tracer of matter distribution, it can be used for mapping the Universe in \emph{Large Scale Structure} (LSS).

However, 21cm emission post-reionization $(z<6)$ is a weak signal which makes its identification hard on a large scale \cite{21cmIM_jeffpeterson_2009}. Thus, measuring each source of emission individually is inefficient. Rather, collecting all emissions in the same beam resolution without distinguishing the sources provides a wider covering in a shorter time. Based on this idea, the BINGO project \cite{BINGO_I}, acronym for \emph{Baryon Acoustic Oscillations from Integrated Neutral Gas
Observations}, aims to map integrated 21cm intensity from the southern hemisphere to be the first to identify Baryonic Acoustic Oscillations (BAO) \cite{2BAOtheoretical_bassett_hlozek_2010} at radio frequency and then to put strong constrain into Dark Sector models. It has been developed to operate at radiofrequency as, firstly, a single-dish with a good resolution and control over the systematic to guarantee the stability of the operation. An approximate prevision for the first pilot tests is for the end of 2023 or the beginning of 2024.

The \emph{Advanced Bingo Dark Universe Studies} (ABDUS) project consists of the next phases of the BINGO project that has already been in studies and should be confirmed completely soon. The new phases aim to explore the current BINGO configuration together with distant stations of feed horns and phased arrays in order to enable the project to achieve, in addition to improving the coverage and survey speed of the current case, a large amount of data from radio sources such as FRBs. BINGO-ABDUS project will be able to cover half of the sky in a quite precise monitoring and overlapping with important other projects in radioastronomy and also galaxy surveys. Such overlap can enable both a good synergy to realize cross-analysis with galaxy datasets and also enable collaboration with other radiotelescopes in order to achieve unbelievable accuracies for individual source identifications. As an example, a collaboration between BINGO-ABDUS, FAST, and Tianlai has the potential to form one of the largest and most accurate radiotelescopes in the world.

This work aims to provide a brief overview of the BINGO project and its current status, and in addition, gives for the first time some information about the BINGO next step to be called BINGO-ABDUS, in which the project will be able to be a versatile and in the state-of-the-art of radio operation for looking for transient radio sources.

\section{The BINGO Project}
The BINGO project consists of a fixed transit telescope covering the south celestial sky by drift scanning mode due to the Earth's rotation. The area coverage is $\sim$ 5000 deg$^2$ between $-25^o$ and $-10^o$ in declination and with an average angular resolution of 40'. The beam resolution in the order of 40' is required to resolve structures of angular scales of $\approx$ 150 Mpc in the BINGO frequency range and then be possible to identify BAO's feature. The telescope will observe signals within a frequency span of 980-1260 MHz (corresponding to a redshift range of 0.127-0.449). 

\begin{figure}
    \centering
    \includegraphics[scale=0.37]{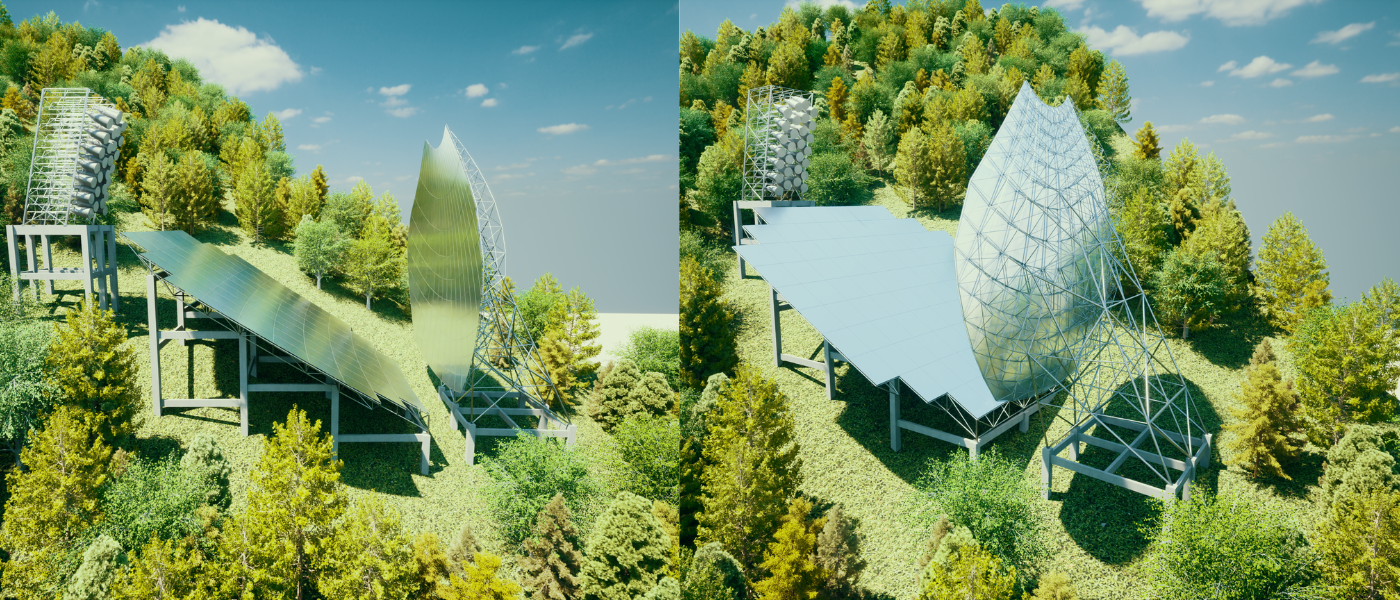}
    \caption{Artistic view of the BINGO's construction model.}
    \label{fig: modelo_prototipo_bingo}
\end{figure}

BINGO has been developed to be an off-axis, crossed-Draconian optical design \cite{dragone1978} composed of a parabolic primary mirror and a hyperbolic secondary mirror with $\sim 40$m diameter, and a complex arrange of 28 vertically moving feed horns on the focal plane, as shown in Figure \ref{fig: modelo_prototipo_bingo}.  Its under-illuminated second mirror guarantees reduced spillover effects and a small amplitude of the far beams' sidelobes. Each receiver operates, so far, with a system temperature of 70 K and measures I and V polarization. In addition, its conical corrugated horns provide low cross-polarization levels as well as a small level of optical aberrations.  A complete description of the instrumentation and the optical design can be found in \cite{BINGO_II, BINGO_III}. 

Currently, BINGO signed a construction contract with the Chinese company CETC54\footnote{54th Research Institute of the China Electronics Technology Group.} to manufacture the reflectors. All the feed horns are already in place and in storage, and there is an additional feed horn (called Uirapuru) that is at the Federal University of Campina Grande (UFCG, in Portuguese acronyms) campus, pointing directly to the sky and doing some data collection.

\begin{figure}
    \centering
    \includegraphics[scale=0.37]{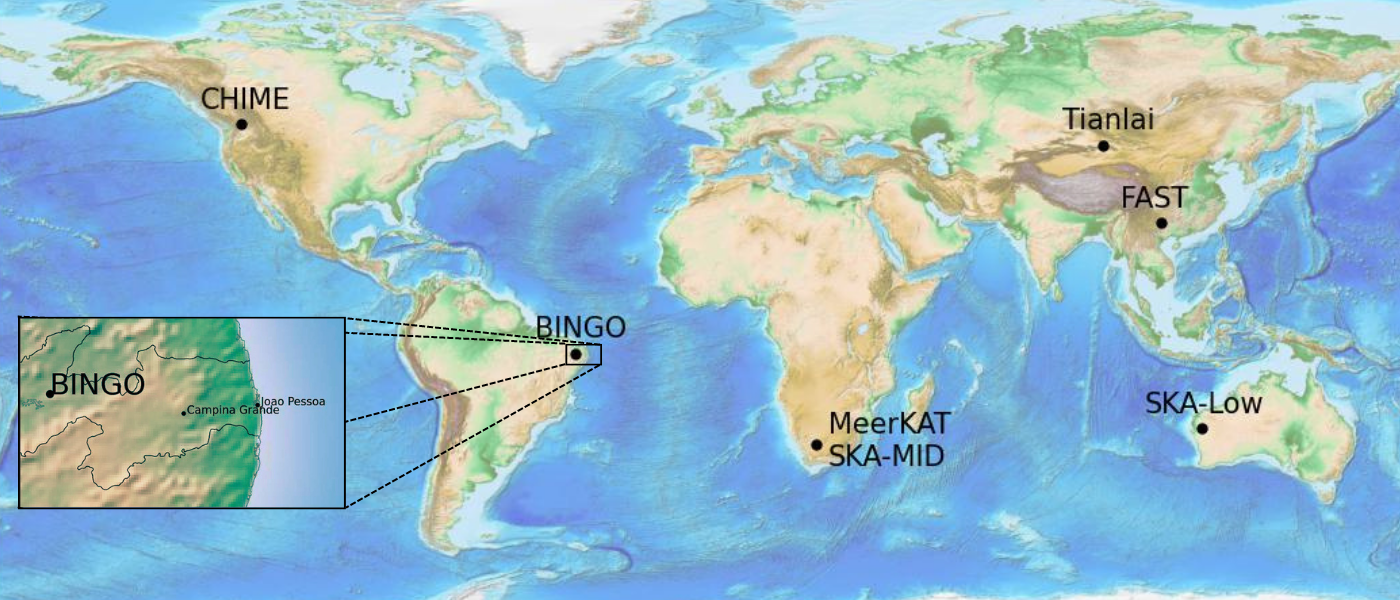}
    \caption{Some HI Intensity mapping projects displayed around the world. There is a focus on the north of Brazil to show the state of Paraiba, and, in its countryside, the BINGO site located in the village of Aguiar.}
    \label{fig: WorldMap_IM_projects}
\end{figure}

The telescope has been constructed in a far site at the village of Aguiar, in the countryside of the Paraiba state. The \emph{Serra do Urubu} site was chosen after a long time investigating different locations since an abandoned gold mine in Uruguay to the northeast of Brazil, as presented in \cite{peel2019}. The choice followed several criteria as \emph{radio frequency interference} (RFI) level, local topography for supporting the telescope structure, and the ease of assistance and follow-up due to the relative proximity to the UFCG.  In Figure \ref{fig: WorldMap_IM_projects} we can see the square focusing on the northeast region of Brazil, where the Paraiba state is located. In this figure, it is possible to see the location of the cost of the country, the capital of the state, and the location of the Campina Grande city.

BINGO telescope aims at using the spatial fluctuations measured from the \emph{Intensity Mapping} observational method to produce tomographic analysis with (auto and cross) angular power spectrum, that is, the statistical two-point function on the harmonic space. The harmonic space is chosen mainly because of the characteristics of observation where there are precise determinations of the redshifts. The tomographic analysis of the two-point functions can be realized in spherical shells.

\subsection{Scientific Organization}
The international BINGO collaboration has concentrated efforts to develop the physical telescope and, at the same time and in parallel, build a complete and well-structured set-up of pipelines to test and cross-check each process of the telescope and its pre- and post-processing data. Such a process will help achieve the best accuracy in data recovery and obtain strong constraints in the cosmological parameters. The collaboration is organized into four work groups, called "stage", each one in charge of covering a set of specific tasks besides the outreach. For sure, there is no precise separation between the starting point and the end for each stage. Indeed, there is a strong superposition among different stages. A brief description of the arrangement of each stage can be seen as follows:
\begin{itemize}
    \item \textbf{Stage 0:} This stage is responsible for handling calibration, tests, and maintenance of the instrumental parts. All processes, from the interaction between the electric field from the sky and the optical design to the analogic-digital conversion data or even the identification of transient astrophysical phenomena are part of the responsibilities. The construction of the huge reflectors has been developed together with CETC54 (China Electronics Technology Group Corporation) company. The development and tests of the feed horns and front-ends (transitions, polarizers, and magic tees) were conducted together with local companies ALLTEC and Metalcard, as presented in \cite{BINGO-testing_horns}. Furthermore, possible \emph{Fast Radio Bursts} (FRBs) \cite{Lorimer_2007, Zhang_2020}, pulsars, and \emph{Radiofrequency Interference} (RFI) are firstly identified in this stage. Machine Learning techniques have been studied to separate chunks of relevant data. FRBs and pulsars phenomena will be monitored in higher precision in the next level of the BINGO, called the ADBUS project, as we will describe soon. A good description of the stage's progress and development can be found in \cite{BINGO_II, BINGO_III}.
    
    \item \textbf{Stage 1:} Stage responsible for the time-ordered data (TOD), map-making, and recording and flagging the possible transient objects. After the data is turned into digital information, it is necessary to record the observed data in a chronological description identifying the main information from the radiometer outputs, pointing locations, and the behavior of the system. Different system effects need to be assumed and understood here in order to achieve precise gain stability and accuracy of the reconstruction of 21cm data. The TODs will be pre-processed into tomographic images corresponding to BINGO covering by the map-making process. Currently, there are two methods for processing TOD and map-making, one from the University of Manchester and another based on the Hide\&Seek\footnote{\small https://cosmology.ethz.ch/research/software-lab/hide---seek.html, https://cosmology.ethz.ch/research/software-lab/hide---seek.html} software that has been extended and adapted to the BINGO case\footnote{A complete description about the Hide\&Seek adaptation has been writing in a new paper to be published soon and presenting newest map-making results and beam modeling from Zernike polynomials \cite{BINGO_HS_zernike}.}.
    One of the main collaboration interests is related to the FRBs, which is a strong astrophysical emission with the origin unknown. A candidate for FRB is not necessarily a reportedly FRB; it is necessary that the candidate fill in a set of criteria to be accepted as an FRB itself. Also, other astrophysical (radio-)transients - such as pulsars - will be monitored by the stage and associated with trained neural networks for identification.
    
    \item \textbf{Stage 2-3:} Firstly separated, stages 2 and 3 have been merged into the same group due to the increasingly overlapping areas of activity. The group is in charge of simulating the process of generating 21cm emissions from N-body and lognormal (distribution) mocks \cite{BINGO_VI}, simulating possible astrophysical and cosmological sources of contamination \cite{BINGO_IV, BINGO_V}, and the process of reconstruction the 21cm signals \cite{BINGO_demericia, marins2022} and BAO identification \cite{BINGO_VIII}. Roughly speaking, we can resume this stage as responsible for the data analysis. 
   
    \item \textbf{Stage 4:} The fourth stage is the most theoretical one, interested in investigating the physics behind the data. Remembering that the BINGO concept has been developed aiming to investigate the Dark Sector of the Universe. A first estimation of the quality in constraint some dark energy - from CPL parametrization and interacting dark energy - and modified gravity parameters, as well as the synergy with other projects, have already been published \cite{BINGO_I, BINGO_VII} forecasting - in a Fisher matrix context of analysis - the BINGO contribution. Other types of physical interests are also taken into account in this stage, such as the study of neutrinos, HI parameter density for a range of the Universe's history, and FRB and Pulsar models. In summary, this stage has the purpose, from the project data, to elucidate and constrain the possible explanations to describe the current Universe dynamics observed. The possible constraint on cosmological parameters can have a large impact on the structure of the Dark Sector, especially concerning an internal interaction between DE and DM. The cosmological consequences of the possible localization of FRBs from the outriggers proposed for the project, especially in the case of the ABDUS project, are expected to be very large. 
\end{itemize}
However, the BINGO project is not only about researching scientific issues but also about how to make the data and information produced available for the (civil and scientific) community and how to help society to be closer to the science and technology developed. BINGO outreach has been focusing on efforts to science communication, building a bridge for continuous collaboration between the expert production from the world-class researchers to the social and educational development. The BINGO outreach activities are being publicized on the BINGO website\footnote{\small {https://bingotelescope.org/}{https://bingotelescope.org/}} and its social media channels.

\section{The ABDUS Project - Advanced Bingo Dark Universe Studies}

\begin{figure}
\centering
    \includegraphics[scale=0.43]{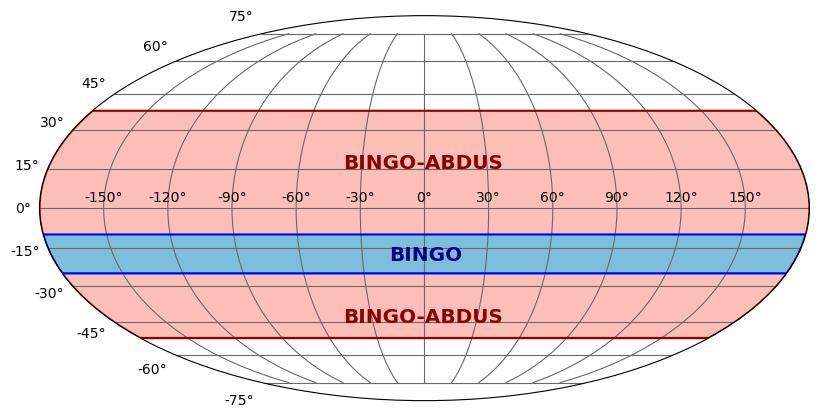}
    \includegraphics[scale=0.43]{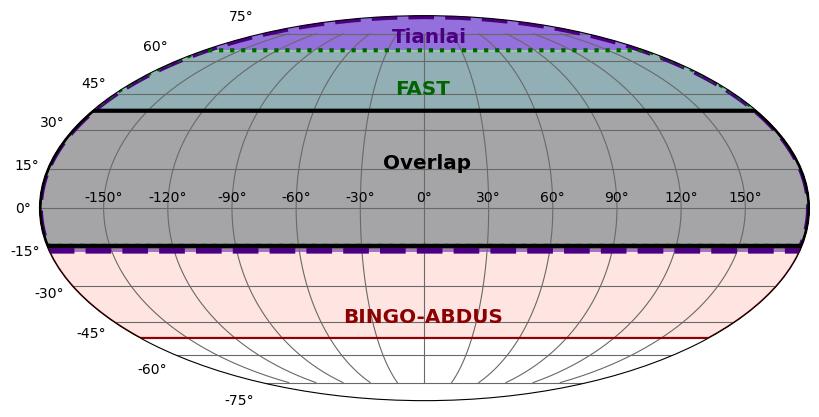}
    \caption{Unmasked coverage by BINGO, ABDUS, FAST, and Tianlai. On the left, it is possible to check the impressive increase in the region to be observed by the BINGO-ABDUS project.  On the right, there is in highlight the Overlap region to be coverage by BINGO, FAST, and Tianlai.}
    \label{fig: overlap_radio_projects}
\end{figure}

The ABDUS project honors the great physicist Abdus Salam, and represents the acronyms for \emph{Advanced Bingo Dark Universe Studies}. It is an ambitious Radio Telescope resulting from a collaboration between BINGO (Brazil and China), the Netherlands Institute for Radio Astronomy (ASTRON), and Nancy Observatory. It will be BINGO's next step towards enhancing its resolution and coverage, and mainly to boost BINGO as one of the main telescopes to search for Fast Radio Bursts and strong radio point sources. For that, ABDUS proposes to use detectors in the form of phased arrays, which will allow expressively enlarging the BINGO's capabilities to reach out to $ z \sim 2.1$, for the 21 IM scanning and covering up to 50\% of the sky (see the left image from Figure \ref{fig: overlap_radio_projects}) in a resolution between 27'-40', and increasing the current number of beams from 28 up to 200 in a multi-beam electronic synthesis through beamformers. This increase will be conducted by joining the already contracted BINGO structure with outrigger and phased array stations. The farthest station will be $\sim 20$ km away from the main telescope site and will make it possible to reach out up to 2.5'' angular resolution to pinpoint individual radio sources by interferometry when they are triggered by the main telescope.

Phased arrays (PAs) are a set of closely distributed antenna elements separated by about half wavelength, which can synthesize multiple independent beams and guarantee (Nyquist's field-sampling) requirement for a uniform sensibility coverage. In contrast to an interferometric approach, PAs work by imaging ''additively'' the PA element outputs, and the number of beams simultaneously synthesized depends on the digital processing capacity. Among the advantages of PAs, we can mention good control over the beam patterns, multiple steered beams, the possibility of achieving continuous and uniform sky coverage, an improvement in the FOV due to the increase of the number of beams leading to a higher survey speed, sensitivity optimization in relation to RFI contamination, and wide frequency coverage.  Different projects use different approaches to build the PA beams, and an important procedure is through a beamforming technique, which is a process of adding all incoming signals (PA element outputs) weighted by appropriate values (beam weights). The illumination pattern is no longer fixed by the optical design but by the beam weights. Different values of beam weights lead to different directions of pointing, and then beamformers provide a way to observe different locations at the same time, which leads to the advantage of observing a calibrator source simultaneously with the target field \cite{beamforming_2022_DCprice}.

The BINGO's today optical configuration is a reflector system with focal-plane arrays (FPAs); the first extension to the ABDUS setup will be associating phased-array to the BINGO focal plane (so-called ABDUS-1), that is, to create a PA receiver system using Vivaldi antennas (single polarized tapered slot antennas) in two different directions, crossing, providing a two orthogonal polarizations and associated to the beamformers. The PAs will take advantage of the current BINGO optical design and will be placed beside the feed horns' structure, with (an initially proposed) total area of $2.8 \text{m} \times 20 \text{m}$, located in a good position of the BINGO's focal plane. At a second moment, there will be outriggers composed of feed horns with mirrors in their mouths and phased-array stations \cite{BINGO_IX} (so-called ABDUS-2). The outriggers and phased array stations will be about 20km distant from the main telescope to provide sufficient angular resolution through an interferometric approach to pinpoint FRBs and other possible radio sources.

The phased array stations are proposed to cover an area of about 36 m$^2$ composed of 32 (sub-) tiles\footnote{A tile is composed by four subtitles of $1.4\times0.35$ m$^2$ arranged parallel to the largest length.} of $1.4\times0.7$ m$^2$. The ABDUS-2 will be able to achieve an impressive leap in the sky coverage. Assuming the latitude of the stations as that of the current BINGO site (lat $7^o 2'27.6''$) and, in a good approximation, the Vivaldi antennas with a great response between $\sim \pm 45^{o}$ in relation to their boresight, the new declination coverage will be within $(-52^o, 38^o)$ and the sky coverage of $32400\text{ deg}^2$. This new sky swath to be covered has a huge overlap with other highlighted radiotelescopes, e.g., $\sim 18846\text{ deg}^2$ considering the FAST and Tianlai region described in the Table \ref{tab: IM_postEoR_projects}, between declination $-14.35^{o}$ and $38^o$, as shown in Figure \ref{fig: overlap_radio_projects}. This provided a fantastic opportunity to operate in a Very Large Baseline Interferometry (VLBI) mode with a huge maximum baseline (see Figure \ref{fig: VLBI_radio_ABDUS_FAST_TIANLAI}).

Several projects are using, or are proposing to use, phased arrays to search Fast Radio Bursts (FRBs), what is crucial due to the low number of current FRB detections ($\sim$ 800); however, though a 21cm IM use was first done exploring the Parkes telescope in a crossed analysis with WiggleZ optical survey between 700-1084 MHz \cite{phasedarray_2021_HI_LinCheng}, ABDUS will open an important window of new and accurate data in a huge sky region, in a greater angular resolution, since the use of phased-arrays in the ABDUS range frequency is a gap to be filled in.  The first analysis and studies address an increase of the Figure of Merit (FOM) for the Dark Energy equation of state from 19 to $(>)$3417, and it can achieve up to 12333 (SKA-MID with band1 and band2 can achieve 244 and 3213, respectively). Considering FRBs, as presented in \cite{BINGO_IX} where we assumed only outriggers with different sizes of mirrors coupling them, for two baselines and 4m mirrors, ABDUS will be able to localize by 20 events per year considering 10 outriggers. In the case of a set of three outriggers in a total of 50 sets (150 outriggers) pointing to the same sky direction, this number leaps to $\sim$180 localized FRBs (by at least 3 baselines) with redshifts up to 1.3.

\section{Other similar projects}
BINGO is not the only project to measure the 21cm post-reionization signals from the Universe. Other projects are operating, under development, or being planned. Each project aims to explore a specific frequency span, with a specific sky covering and strategy of observation, developing new instrumental tools and data treatment to achieve the data as precisely as possible within its own propositions. Even though the road to data would be slightly different, the overall goal is the same: helping the scientific community to unveil the nature of the Universe's dynamic behavior and not well-understood astrophysical phenomena.

Considering mainly the post-reionization era, and just to mention some projects, we give below a small summary. Not necessarily these projects below are constrained to the post-reionization, but they have good synergy with the BINGO goals, and all projects to be mentioned are complementary.

\begin{itemize}   
    \item \textbf{CHIME:} The Canadian Hydrogen Intensity Mapping Experiment (CHIME) is a radio interferometer located at the Dominion Radio Astrophysical Observatory (DRAO), near Penticton, British Columbia province of Canada, operating as a fixed and transient telescope exploring the drift scan mode covering all northeast celestial sky with a field-of-view (FOV) of $\sim 100^o$ in declination. The experiment consists of an array with four 20m $\times$ 100m cylindrical telescopes - the shorter dimension has a parabolic curvature -, each instrumented with 256 dual-polarized feeds \cite{chime_2022_amiri_overview}. The angular resolution is approximately 40' and the frequency resolution is 390 kHz spanning 400-800 MHz (corresponding 21cm emissions from redshift range 0.8-2.5). In 2022, the CHIME collaboration reported the first detection of a 21cm signal in cross-correlation with galaxies (luminous red galaxies and emission-line galaxies) and quasars by the eBOSS catalogs \cite{eBOSS_2016} through intensity mapping measurements made with an interferometer telescope.
    
   \item \textbf{MeerKAT:} located in the arid Upper Karoo region, Northern Cape Province of South Africa, the MeerKAT\footnote{MeerKAT name has double meaning: ''more of KAT'',  where ''meer'' is the Afrikaans word for ''more'' and KAT is the acronyms for ''Karoo Array Telescope''; and, at the same time, it is a name of a suricate, a member of the mongoose family indigenous to Southern Africa.} radio telescope is an array composed of 64 dishes of 13.5m diameter - each one being an offset Gregorian design - over an area with a diameter of 8 km. 48 dishes are placed within 1 km diameter, and the other 16 antennas are mounted at larger distances up to a 4 km radius from the center. The radio telescope can operate using three possible receivers spanning different frequency bands: UHF-band ($580-1015$ MHz), L-band ($900-1670$ MHz), and S-band ($1750-3500$ MHz). MeerKAT, which will become a part of the final SKA1-Mid (see more above), can operate in either interferometric or single-dish mode. Single-dish observations were carried out with L-band (achieving z $< 0.56$), first in a pilot survey data covering a limited celestial region $153^o <$ RA $< 172^o$ and -$1^o <$ Dec $< 8^o$ \cite{meerkat_2021_wang}; later, reporting detection of cross-power spectrum \cite{meerkat_2023_cunnington} with WiggleZ data \cite{wiggleZ_2010} and the first direct detection of the HI intensity (auto-) power spectrum through intensity mapping at $z\approx 0.32$ and $z\approx 0.44$ \cite{meerkat_2023_paul}.
    
    \item \textbf{SKA:} The conception of the \emph{Square Kilometre Array} (SKA) radio telescope dates back to the 1990s - even without this name - with scientists and engineers from different parts of the world joining efforts to design and build the most sensitive radio telescope to date and only possible to be financed in cooperation on a global scale. The name \emph{Square Kilometre Array} (SKA) was assumed in a radio astronomy conference in Calgary, province of Alberta, Canada, in 1998 \cite{ska_2012_history_ekers}, combining scientific and technical development to investigate the Universe in a large range of its history. Currently, the radio telescope is the biggest planned radio telescope to be operating in the Intensity Mapping observational method. 
    
    The SKA's 21cm IM survey operation is split into two phases (SKA1 and SKA2), but here we are commenting only on the most developed one, which is the SKA1\footnote{The SKA2 (also named as SKA full) configuration has not been decided yet; it has only been traced the main scientific and technical goals \cite{ska_2015_braun}, achieving a sensibility 10x and a field of view 20x higher than the first phase.}. The SKA phase 1 (SKA1) can be separated into two parts due to the receiver frequencies, location, and observational mode: SKA1-LOW and SKA1-MID. 
    
    SKA-Low is hosted in the shire of Murchison, 300km northeast of Geraldton,  Western Australia, located on the traditional lands of the Wajarri Yamaji (Wajarri people). The area comprises 512 stations, with 256 dipole antennas per station, operating in an interferometric mode with a maximum distance between antenna stations of 74km. Each station has its antennas distributed out within an area of a 35m radius - designed to support a FOV of $\sim 20\ \text{deg}^2$ at $\sim 100 \text{MHz}$ -, and the core stations - composed of 224 stations and comprising most of the large-scale sensitive - are spread over 500m, arriving at a maximum baseline size of 1km \cite{ska_2020_redbook_dbacon}. This SKA part is prioritized for imaging and spectral observations of the 21cm emissions at high redshift. Its stations will be observing in a 35m (beam former) station FWHM of $\sim 5^o$ (at 100 MHz) and in a frequency range 50-350 MHz (corresponding to $z$ in the interval $3.06 $ to z $\sim$ 27.4, that is, crossing the epoch of reionization and reaching around the Cosmic Dawn epoch when the first lights arose), but only 50-200 MHz (for $z$ in the interval 4.91 to 27.4) will be used to 21cm signals \cite{ska_2017_ska-low_labate}.

    \begin{figure}
\centering
    \includegraphics[scale=0.6]{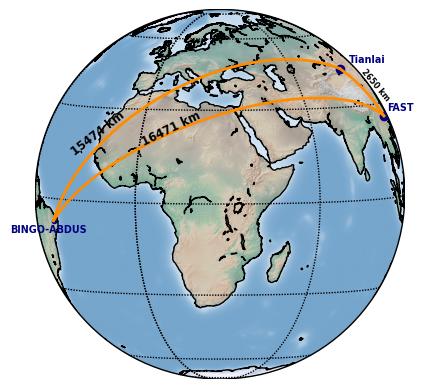}
    \caption{Triangle formed by BINGO-ABDUS, FAST, and Tianlai and the distance between each on of them in kilometers. }
    \label{fig: VLBI_radio_ABDUS_FAST_TIANLAI}
\end{figure}

SKA-MID will be located at the same site as the MeerKAT (Karoo semi-desert plateau)  once the latter is an SKA precursor. Its setup is composed of 64 13.5m dishes from MeerKAT plus 133 (Gregorian offset) dishes with 15m diameter, displayed in a compact center of 1km diameter followed by a 3-arm spiral configuration reaching 150km in the distance between the two extreme dishes.  The frequency range is determined through five receiver bands: Band1 350-1050 MHz, Band2 950-1750 MHz, Band3 1650-3000 MHz, Band4 3000-5200 MHz, and 4600-15800 MHz\footnote{An additional Band6 is desired to cover the frequency range 15-24 GHz.}, with only Band1 and 2 being used to the 21cm signal observations for redshifts $\leq 3$ and a covering of $\sim 30000\ \text{deg}^2$ (in a few thousand hours). The maximum baseline, 150km, at 1.4GHz, corresponds to an angular resolution $\sim 0.3'$. SKA-MID has the interesting issue of being able to work in both interferometric and single-dish modes. For the single-dish case, each dish can independently operate, and the resulting maps are co-added.

   \item \textbf{FAST:} the \emph{Five-hundred-metre Aperture Spherical Telescope} is the largest and most sensitive (filled-aperture) single-dish radio telescope in the world \cite{FASTNan}. The telescope is located in a karst depression called Dawodang in the southern of the Guizhou province, southwest of China. The main FAST reflector is a 500m paraboloid with an illumination aperture of 300m in diameter. The telescope reaches the impressive milestone of an effective collection area about 1.6 times larger than that of the Arecibo telescope \cite{arecibo_2002}. FAST reaches,  at 1.6 GHz, a beam size of 2.9' (3.5' of Arecibo), a system temperature of 18 K (30 K, Arecibo), spanning a frequency range of 70 - 3000 MHz, with seven sets of receivers, for $-14.4^o <$ Dec $< 65.6^o$ \cite{fast_2020_outreach_qian}. For 21cm IM observations, there is an L-band 19-beam receiver between 1.05-1.45 GHz, with 18 feeds arranged in two concentric hexagonal rings plus one feed in the center. In addition, thanks to its high angular resolution, it is expected to resolve a large number of galaxy clusters, groups, filaments, and voids, reaching the detection of until $6.5 \times 10^5$ HI galaxies for $z\sim 0.07$ and a mass threshold of $10^{9.5}M_{\odot}$. Commissioning and pilot tests were done for the 21cm IM drift-scan mode (with the L-band 19-beam feed system) in 2019, 2020, and 2021 to investigate frequency resolution to be assumed to allow a low level of radio frequency interference (RFI) and 1/f effects, level of the noise diode power, rotation of the feed arrangement to achieve another declination regions in the sky \cite{fast_2021_whu_1f, fast_2021bzhang_HIabsorption_lines_obs, fast_2023_YLi_preliminary_datanalysis}.
    
    \item \textbf{Tianlai:} This name in Mandarin means ''heavenly sound'' and represents here the international but Chinese-leaded project of 21cm signals post-reionization through interferometry Intensity Mapping. Located at Hongliuxia village, Balikun county, in the Xinjiang region, northwest China, the experiment covers the entire northern celestial sky (and also part of the southern one) using the Earth's rotation \cite{tianlai_2018_Das_construction}. At present, the project consists of two radio-interferometers: one is composed of three arrays of adjustable parabolic cylinder reflectors with 40m $\times$ 15m size\footnote{The goal is the full-scale Tianlai experiment consist of eight parallel cylinders increasing the length from 40m to 120m, with a total of 2000 dual-polarization feeds in a frequency range spread between 400-1420 MHz \cite{tianlai_2015_xu}.}, composed of 96 dual-polarization feeds, and oriented in the N-S direction (called \emph{Tianlai Cylinder Pathfinder Array}). The other radio-interference system (called \emph{Tianlai Dish Pathfinder Array}) is arranged in a near-hexagonal configuration composed of 16 on-axis dishes of 6m aperture and disposed of in two concentric rings \cite{tianlai_2021_Wu_design}. 
    At this first stage, called Pathfinder, the telescope operates at 700-800 MHz (corresponding to the redshift range 0.78-1.03). The future observations are planned to achieve the frequency range 1330-1430 MHz (for 21cm emissions, this corresponds to the redshifts $<0.068$).    
\end{itemize}

\begin{table}[h]
\footnotesize
\centering
\caption{All projects are described here following your 21-cm intensity mapping operation. We will not focus on any other operational frequency bands or suitable characteristics in source monitoring and individual localizing. Any project can possibly have new features not covered here.}
\begin{tabular}{l c c c c c c c c }
 \hline
Project & Type & $\theta_{\footnotesize \text{FWHM}}$ & Freq. (MHz) & N. beams & Cover (deg$^2$) & Dec.\footnote{Declination.} (deg) & $\text{T}_{\footnotesize \text{sys}}$ (K)\\
\hline
BINGO & Single-dish & 40' & 980-1260 & 28 & 5324 & (-25,-10) & 70\\ 
ABDUS-1 & Phased-array& 40' & 450-1420 & 44 & 5324 & (-25,-10) & 40-60\\ 
ABDUS-2& Phased-array& 27' & 450-1420 & 200 & 32400 & (-52,38) & 40-60\\ 
Tianlai Cyl.\footnote{Value based on \cite{tianlai_2020_systemfunction}.} & Cyl. interf. &   1.8$^o$\footnote{This value assumes the response for the E-W scan direction. For the N-S direction, the cylinder is basically unfocused with >60$^0$ \cite{tianlai_2020_systemfunction}. Both measures mentioned are assumed at 750 MHz.}  & 700-800 & 96 & 10000 & (-16,90) &   82-100\\ 
Tianlai Dish\footnote{Data based on \cite{tianlai_2021_Wu_design, tianlai_2022_forecast}.} & Dish interf. & 12' & 700-800 & 16 & 1500\footnote{Coverage of 1500 deg$^2$ after masking. The declination range covers can be tuned to another sky region due to the mobility of the dishes.} & (48,60) &  80\\ 
FAST\footnote{CRAFTS. Values extracted from \cite{fast_2020_whu_forecast, fast_2021_yohana}.}  & Single-dish &   3' & 1050-1450 & 19 & 20000 & (-14, 66) & 20\\ 
CHIME\footnote{Data based on \cite{chime_2022_amiri_overview}.} & Cyl. interf. & 40' & 400-800 & 1024 & 31000 & (-21, 65) & 55\\ 
MeerKAT\footnote{We are assuming in this table the values provided for L-band, extracted from \cite{meerkat_2021_wang, meerkat_2023_cunnington}. MeerKAT will also be able to operate in other modes, and one of the ongoing observations is using a UHF band covering 10000 deg$^2$. } & Single-dish & 1.0$^o$-1.6$^o$ & 900-1420 & 64 & 3000 & (-1, 8) & 18-22\\ 
SKA1-MID\footnote{Values assuming only Band1 in an autocorrelation mode as \cite{ska_2020_redbook_dbacon}. The declination corresponding to the coverage mentioned in the main papers was not identified. } & Single-dish & 30'-4$^o$ & 350-1050 & 64+133 & 30000 & - & 25-63\\ 
\hline
\end{tabular}
\label{tab: IM_postEoR_projects}
\end{table}


All the above-cited projects pursue not only 21cm signals as a scientific goal. They are also important in the investigation of transient phenomena such as pulsars and fast radio bursts (FRB).

\bibliographystyle{unsrt}
\bibliography{main}

\end{document}